\documentclass[%
 aip,
 sd,%
 amsmath,amssymb,
 reprint,%
]{revtex4-1}

\usepackage{graphicx}
\usepackage{dcolumn}
\usepackage{bm}

\begin{document}

\preprint{AIP/123-QED}

\title[Coherent structure coloring]{Identification of individual coherent sets associated with flow trajectories using Coherent Structure Coloring}

\author{Kristy L. Schlueter-Kuck}
 \affiliation{Department of Mechanical Engineering, Stanford University.}
 
\author{John O. Dabiri}
 \email{jodabiri@stanford.edu}
 \affiliation{Department of Mechanical Engineering, Stanford University.}
\affiliation{ 
Department of Civil and Environmental Engineering, Stanford University.
}%

\date{\today}

\begin{abstract}
We present a method for identifying the coherent structures associated with individual Lagrangian flow trajectories even where only sparse particle trajectory data is available.    The method, based on techniques in spectral graph theory, uses the Coherent Structure Coloring vector and associated eigenvectors to analyze the distance in higher-dimensional eigenspace between a selected reference trajectory and other tracer trajectories in the flow.  By analyzing this distance metric in a hierarchical clustering,  the coherent structure of which the reference particle is a member can be identified.  This algorithm is proven successful in identifying coherent structures of varying complexities in canonical unsteady flows.  Additionally, the method is able to assess the relative coherence of the associated structure in comparison to the surrounding flow.  Although the method is demonstrated here in the context of fluid flow kinematics, the generality of the approach allows for its potential application to other unsupervised clustering problems in dynamical systems such as neuronal activity, gene expression, or social networks.
\end{abstract}

\maketitle

\begin{quotation}
In recent years, there has been a proliferation of techniques that aim to characterize fluid flow kinematics on the basis of Lagrangian trajectories of collections of tracer particles. Most of these techniques depend on presence of tracer particles that are initially closely-spaced, in order to compute local gradients of their trajectories. In many applications, the requirement of close tracer spacing cannot be satisfied, especially when the tracers are naturally occurring and their distribution is dictated by the underlying flow. Moreover, current methods often focus on determination of the boundaries of coherent sets, whereas in practice it is often valuable to identify the complete set of trajectories that are coherent with an individual trajectory of interest. We extend the concept of Coherent Structure Coloring, an approach based on spectral graph theory, to achieve identification of the coherent set associated with individual Lagrangian trajectories. The method does not require a priori determination of the number of coherent structures in the flow, nor does it require heuristics regarding the eigenvalue spectrum corresponding to the generalized eigenvalue problem. Importantly, although the method is demonstrated here in the context of fluid flow kinematics, the generality of the approach allows for its potential application to other unsupervised clustering problems in dynamical systems such as neuronal activity, gene expression, or social networks. 
\end{quotation}


\section{\label{sec:intro}Introduction}
The ability to detect regions of coherence in fluid flows is of interest to many scientific communities.  For example, by identifying groups of fluid particles that move coherently, it may be possible to characterize how passive scalars like heat and salt concentration are transported by currents and eddies in the ocean~\citep{Abernathey2017}.  Because of the underlying interest in transport phenomena, objective Lagrangian methods, which are independent of the reference frame, are a logical tool for addressing these questions.  Lagrangian methods for coherent structure detection have been studied intensively over the last several decades (see~\citet{Hadjighasem2017, Allshouse2015} for reviews), especially in flow fields that are well characterized.  If full-field velocity data is available, deformation gradient-based techniques are useful.  For example, calculation of the finite time Lyapunov exponent (FTLE) field can be used to identify the boundaries of fluid regions that experience minimal mixing with the surrounding fluid~\citep{Haller2000,Shadden2005}.

However, in many flows of interest, full field velocity data is unavailable, and we must rely on the advection of a relatively sparse set of Lagrangian particles to characterize the flow.  This occurs because of an inability to densely seed the region of interest of the flow, e.g. in oceanic environments where GPS-enabled drifters or natural seeding is necessary; and in very large volumetric flow domains to measure naturally occurring atmospheric flows.  In these cases, the assumptions inherent in the calculation of the deformation gradient are not valid, especially the ansatz of initially closely-spaced trajectories.  In these cases, other techniques must be used to accurately detect regions of fluid coherence~\citep{Froyland2015, Allshouse2012}.

Recently, several approaches to spectral clustering have emerged to address the problem of coherent structure identification.  In general, these approaches involve building an adjacency matrix containing information regarding the similarity of every pair of particle trajectories.  Eigendecomposition is then used to sort the trajectories according to their similarity, with the sorting information that is contained in specific eigenvectors dependent on the particular method.  The spectral clustering technique of ~\citet{Hadjighasem2016} quantifies the similarity of trajectory pairs based on the mean displacement between the trajectories.  Subsequently, the eigenvectors corresponding to the eigenvalues below a presumed spectral gap, i.e. the largest gap in the value between two adjacent eigenvalues, separate the coherent structures in the flow from the presumed incoherent background.  Each coherent structure corresponds to a separate eigenvector.  This method allows for detection of an arbitrary number of structures in the flow, unlike fuzzy c-means clustering~\citep{Froyland2015}, but is dependent on the eigengap heuristic to determine the number of structures.  The location of this gap can be extremely sensitive to the number of particles tracked, the initial location of the particles in the flow field, and any applied sparsification of the adjacency matrix.  Another recently developed spectral graph theory method~\citep{Schlueter-Kuck2017}, quantifies the relationship between particle trajectories by their kinematic dissimilarity, a weighted measure of the standard deviation of the distance between the two trajectories.  This method, which allows for an analog spectrum of coherence instead of a binary distinction between coherent structures and an incoherent background flow, displays the most significant dissimilarities between particle trajectories in the coherent structure coloring (CSC) field.

The CSC method differs from other graph theoretical methods in the atypical definition of coherence as kinematic similarity of Lagrangian trajectories regardless of spatial proximity.  This unique definition leads to corresponding changes to the conventional adjacency matrix and its analysis, as described in this paper.  The more versatile definition of coherence renders the problem of identifying all of the coherent structures in the flow as ill-posed, making the present method distinct from, and complementary to, other coherent structure identification algorithms~\citep{Banisch2017,Hadjighasem2016,Froyland2015,Froyland2017,Allshouse2012}.  We show that it is generally more useful to consider an individual Lagrangian particle trajectory, and to identify the set of other trajectories that are coherent with the reference trajectory of interest. 

There are several potential advantages of the CSC method, as it does not require determination of the number of coherent structures in the flow, either a priori or through the use of an eigengap heuristic.  Also, it facilitates the identification of coherence among trajectories that remain spatially separated (e.g. regions that are not simply connected), as long as their trajectories are kinematically similar.  However,~\citet{Schlueter-Kuck2017} allude only briefly to potential methods for extracting individual structures from the CSC field. 

This paper describes a method that uses the CSC field and additional information from other eigenvectors of the spectral decomposition to isolate specific flow structures associated with individual flow trajectories.

\section{\label{sec:methods}Methods}

In the CSC algorithm~\citep{Schlueter-Kuck2017}, dissimilarity between two particle trajectories is represented numerically using a weighted adjacency matrix $A$, where $a_{ij}$ contains the weight of the edge connecting particle $i$ and particle $j$:
\begin{equation} \label{eq:adjacency}
a_{ij}=\frac{1}{\overline{r_{ij}}T^{1/2}}\left[\sum_{k=0}^{T-1}(\overline{r_{ij}}-r_{ij}(t_k))^2\right]^{1/2}
\end{equation}
where $r_{ij}(t_k)$ is the distance between two particles $i$ and $j$ at time $t_k$, and $\overline{r_{ij}}$ is the average distance between the two fluid particle trajectories.  Conceptually, $a_{ij}$ quantifies the standard deviation of the distance between particle trajectories normalized by their average spacing.  The corresponding eigenvalue problem that computes the difference between dissimilar particles is 
\begin{equation} \label{eq:gen_eig}
LX=\lambda DX
\end{equation}
where 
\begin{equation}
d_{ij}= \left\{
\begin{array}{ll}
    0,			& i\neq j\\
    \sum_{k=1}^N a_{ik},       & i=j.
\end{array} \right.
\end{equation}
and $L=D-A$ is the graph Laplacian.  In order to maximize the differences between dissimilar particles, $X_1=X$ is the eigenvector associated with the maximum eigenvalue, $\lambda_1$ of this problem, under the constraint that $X^\prime DX$ remains finite.  Each element of $X_1$ assigns that value of CSC to the corresponding fluid particle at the final time of the interval over which particle trajectories were compared.

While the CSC field highlights the largest dissimilarities in the flow, other dissimilarities are also present, information about which is contained in the eigenvectors associated with the eigenvalues less than $\lambda_1$.  We denote these eigenvalues and eigenvectors, $\lambda_1>\lambda_2>\lambda_3...$ and ${X_1, X_2, X_3}$, respectively.  It should be noted that because $A$ is real and symmetric, all $\lambda$ are real, and all $X$ are orthogonal.  Due to these properties, the eigenvectors associated with lesser eigenvalues may contain additional, unique information regarding how to partition the flow to separate dissimilar trajectories.    The quadratic placement algorithm of~\citet{Hall1970} suggests that the optimal distribution of trajectories in an $n$-dimensional space to separate dissimilar trajectories is given by the eigenvectors associated with the $n$ largest eigenvalues of the generalized eigenvalue problem given by equation~\ref{eq:gen_eig}.

One way to partition the flow is to examine the distance from every particle in the flow to a selected reference trajectory in CSC space, or $d_1(i)=|CSC(i)-CSC(ref)|$.  More generally, it is also possible to create a weighted distance metric $d_{wD}$, in an arbitrary number of dimensions $D$, where each element in the distance metric is weighted by a factor $w_{d}$ associated with the corresponding eigenvalue.  Formally,
\begin{equation} \label{eq:dist_metric}
d_{wD}(i)=\left(\sum_{d=1}^{D}w_{d}(X_d(i)-X_d(ref))^2\right)^{0.5}
\end{equation}
This reduces to the unweighted distance metric when $w_{d}=1$ for all $d$.  The distance metric is analogous to the diffusion distance of ~\citet{Coifman2006} in that it is a representation of the connectivity of a pair of fluid particles.  For comparison to the unweighted distance metric, we have chosen
\begin{equation} \label{eq:dist_weight}
w_{d}=\left((\lambda_d-\lambda_{D+1})/(\lambda_1-\lambda_{D+1})\right)^{0.5}
\end{equation}
in order to weight eigenvectors with larger corresponding eigenvalues more strongly.  Different weights can be used without loss of generality, and in this analysis we have found that the unweighted distance metric is effective in isolating individual coherent structures.

Smaller eigenvalues correspond to less effective solutions to the maximization problem given by equation~\ref{eq:gen_eig}; the corresponding eigenvectors contain less and less useful information about flow dissimilarities.  Therefore, the distance metric fields corresponding to a specified reference particle (e.g distance contour plots generated by interpolation of the eigenspace distances) will reach a plateau, where increasing the dimensionality of the eigenspace does not significantly change the distance metric field.  Once this plateau is observed, any smaller eigenvalues and corresponding eigenvectors are relatively unimportant and can be disregarded.

Once the critical dimensionality of the distance metric space of interest is determined (i.e. the aforementioned plateau in eigenspace distance versus number of eigenvalues), a threshold  eigenspace distance can be identified within which the reference particle is coherent with other particles in the flow.  This threshold is identified by using hierarchical clustering to separate the distance metric field into particles within the coherent structure and those outside of it.  Hierarchical clustering begins by considering every particle a distinct group and combining the two groups with the most similar associated distance metric.  The larger group is then assigned a distance metric value that is the average of the values of the two particles in it, and the next most similar pair of particles (or particle groups) is subsequently combined.  This process is repeated until there are only two groups remaining, corresponding to the particles within the coherent structure associated with the selected reference point and those outside of it.

The application of this algorithm to identify specific coherent structures is described in the following section.

\section{Results}
The effectiveness of the aforementioned algorithm for extracting structures associated with individual flow trajectories from the CSC field is demonstrated using two example flows.  The unsteady quadruple gyre flow is used to illustrate the application of this method to detect both vortex cores and the secondary structures caused by oscillating fluid.  A second flow, the Bickley jet, shows that the algorithm is capable of detecting vortical structures as well as elongated jet-like structures in the flow, both of which experience little mixing with the surrounding fluid, and therefore contribute to fluid transport, but have significantly different shapes.  This flow is also used to demonstrate the method's treatment of incoherent background flow.  

\subsection{Quadruple Gyre}
First, we examine the characteristics of the CSC algorithm using the analytical quadruple gyre flow.  This flow is defined by
\begin{eqnarray}
\frac{dx}{dt} & =&-\pi A\sin(\pi f)\cos(\pi y)\\
\frac{dy}{dt} & =&-\pi A\cos(\pi f)\cos(\pi y)(2ax+b)
\end{eqnarray}
where $x$ and $y$ are the spatial coordinates, $t$ is time, and
\begin{eqnarray}
a =\epsilon\sin(\omega t), 
b =1-2\epsilon\sin(\omega t), 
f =ax^2+bx.
\end{eqnarray}

\begin{figure*}
\centering
\includegraphics[width=\linewidth]{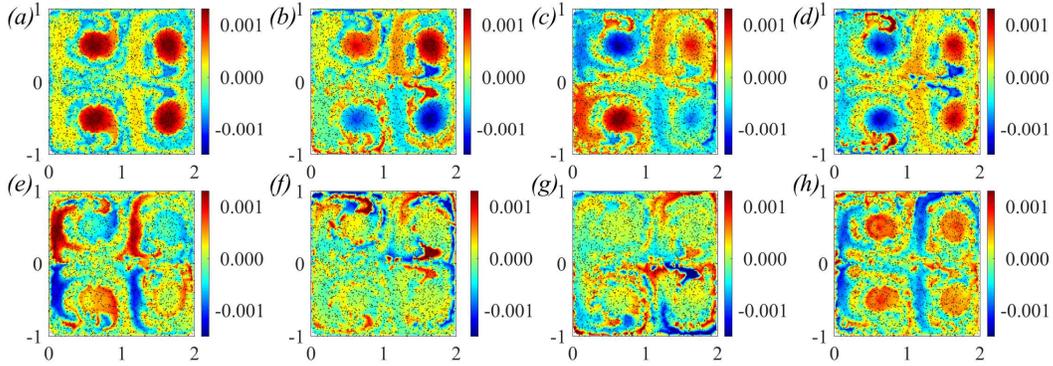}
\caption{Unsteady quadruple gyre flow, $\epsilon=0.1$, $A=0.1$, and $\omega=2\pi/10$, calculated over the time interval $T$=[2.5, 42.5], using 3000 particles.  The eight eigenvectors, $X_1$-$X_8$, associated with the eight largest eigenvalues, $\lambda_1$-$\lambda_8$ of $LX=\lambda DX$ of the unsteady quadruple gyre flow.  Black dots show final location of 3000 particles.  a) the CSC field, $X_1$. b) $X_2$. c) $X_3$. d) $X_4$. e) $X_5$.  f) $X_6$. g) $X_7$. h) $X_8$. } 
\label{fig:gyre1}
\end{figure*}

We consider the unsteady case where $A=0.1$, $\epsilon=0.1$, and $\omega=2\pi/10$.  3000 particles were artificially seeded in the domain and advected with the flow.  Further details of this flow can be seen in~\citet{Schlueter-Kuck2017}. Figure~\ref{fig:gyre1} shows the eigenvectors associated with the largest eight eigenvalues of the eigenvalue problem $LX=\lambda DX$.  The contour plot constructed from instantaneous particle locations and the corresponding eigenvector associated with the largest eigenvalue is the CSC field (figure~\ref{fig:gyre1}(a)).  It is clear that the CSC field and $X_2$ through $X_5$ contain information related to the coherent structures we expect to see in the flow, namely the gyre cores and the secondary structures to the left of each core, while $X_6$ through $X_8$ appear progressively noisier and contain less useful information. 

It is also interesting to note that there is not one structure associated with each eigenvector, as with other spectral graph-theoretic methods~\citep{Hadjighasem2016}.  This is because large edge weights in the adjacency matrix correspond to pairs of particles that are very dissimilar, instead of very similar.  Additionally, because the edge weights depend on both the standard deviation of the particle separation and the magnitude of the separation, small weights are not necessarily associated with particles in close proximity.  This means that the solution to the generalized eigenvalue problem is not an attempt to solve an N-cut problem, as posed by~\citet{Shi2000} and implemented by~\citet{Hadjighasem2016} to partition fluid flows.  Furthermore, an examination of the eight eigenvectors shown here shows that a particle in what is commonly considered a coherent structure (e.g. the upper left vortex core) is not isolated from the rest of the flow in any of the eight eigenvectors shown, but can be isolated by considering that all of the particles in that structure have similar values in \textit{all eight} eigenvectors and different values from the rest of the particles in \textit{at least one} of the eigenvectors.

\begin{figure*}
\centering
\includegraphics[width=\linewidth]{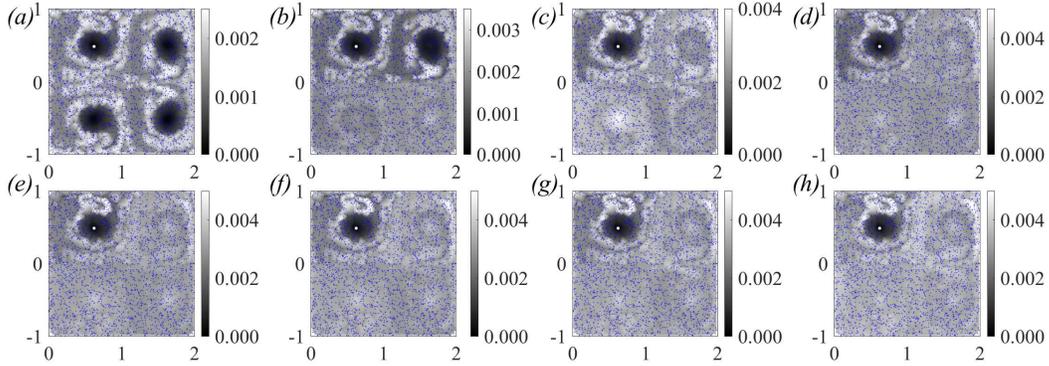}
\caption{The unweighted distance, in eigenspace, to the reference particle indicated by the white dot.  Blue dots represent final locations of all other particles in the flow.  a) one-dimensional eigenspace ($D=1$).  b) two-dimensional eigenspace ($D=2$). c) three-dimensional eigenspace ($D=3$). d) four-dimensional eigenspace ($D=4$). e) five-dimensional eigenspace ($D=5$). f) six-dimensional eigenspace ($D=6$). g) seven-dimensional eigenspace ($D=7$). h) eight-dimensional eigenspace ($D=8$).} 
\label{fig:gyre2}
\end{figure*}

We will consider first a reference particle located in the center of the upper left gyre core, close to the location of the local maximum in the CSC field.  Using the distance metric given in equation~\ref{eq:dist_metric} and a weighting factor $w_d=1$, the distance metric relative to that reference point is calculated for $D=1$ through $8$.  The resulting interpolated distance metric fields are shown in figure~\ref{fig:gyre2}, where the reference particle is indicated by the white dot.  
It is evident that as the dimensionality of the eigenspace is increased, the other particles in the structure containing the reference point, in this case the counterclockwise rotating vortex in the upper left corner of the domain, retain a small eigen-distance from the reference point, while particles in other structures, including those with a similar CSC value to the reference point; i.e. the other three gyre cores, become increasing far away in eigenspace.  As the dimensionality of the eigenspace approaches eight, the distance-metric field changes very little as the dimensionality is increased.  

\begin{figure}
\centering
\includegraphics[width=0.75\linewidth]{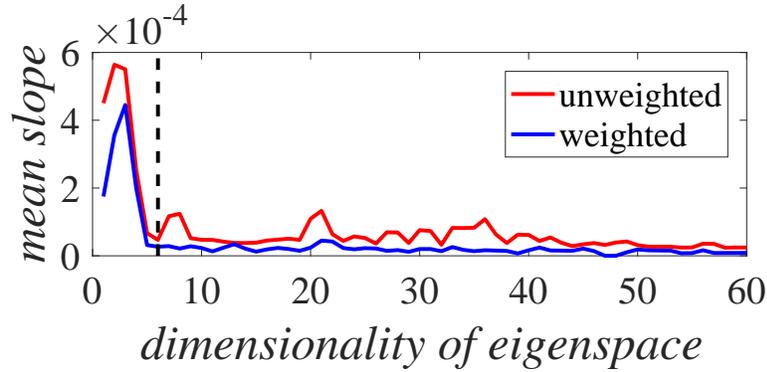}
\caption{Analysis of a reference particle in the vortex core of the upper left quadrant of the quadruple gyre flow: the average slope of the distance to the reference particle ($s(D)$) vs. the dimensionality of the eigenspace ($D$).  Red line is calculated using the unweighted distance metric, where $w_d=1$.  Blue line is calculated using the weighted distance metric, where $w_d$ is given by equation~\ref{eq:dist_weight}.  Vertical dotted black line indicates the location of the elbow in the plot, selected for subsequent analysis ($D=6$).}
\label{fig:gyre3}
\end{figure}

We can quantify the eigenspace dimension at which further increasing the dimensionality of the eigenspace no longer causes significant changes in the distance metric field, by plotting the average slope of the distance metric for every particle vs. the dimensionality of the eigenspace, given by the equation
\begin{equation}
s(D)=\frac{1}{N}\sum_{i=1}^Nd_{wD}(i)-d_{w(D+1)}(i)
\end{equation}
where $N$ is the total number of particles tracked.  This function is shown in figure~\ref{fig:gyre3} for both the unweighted distance metric ($w_{d}=1$ for all $d$) in red, and the weighted distance metric, where $w_{d}$  is given by equation~\ref{eq:dist_weight} in blue.  It is clear that beyond a dimensionality of 6, the contributions to the unweighted distance metric become negligible.  This approach is similar to finding the elbow in a scree plot for principal component analysis~\citep{James2013}.  We have found that the resulting coherent structure identified using this method is robust to the chosen dimensionality of the eigenspace.

\begin{figure}
\centering
\includegraphics[width=\linewidth]{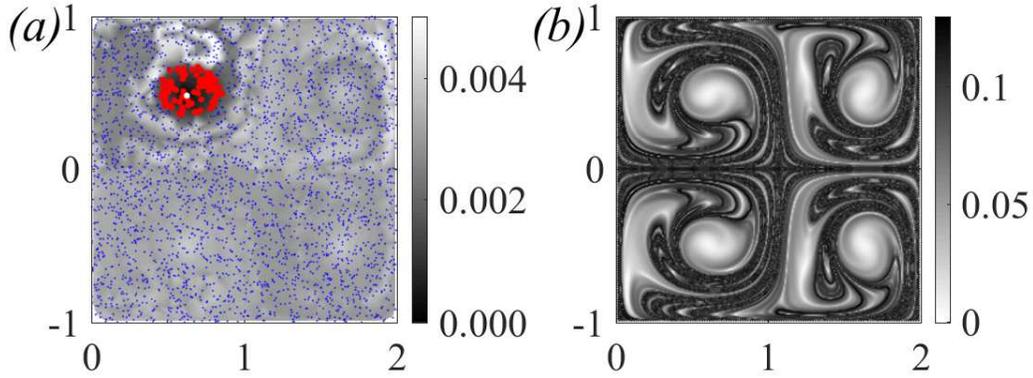}
\caption{a) Analysis of a reference particle in the vortex core of the upper left quadrant of the quadruple gyre flow.  Dots indicate final particle positions: white dot indicates the chosen reference point, red dots are the particles that have been identified as within the coherent structure associated with the reference point, and blue dots  indicate particles outside of that structure.  The underlying field is the 6D unweighted distance metric field. b) FTLE field calculated using 65,000 particles} 
\label{fig:gyre4}
\end{figure}

In order to determine the separation between the particles inside the same structure as the reference point from those outside, agglomerative hierarchical clustering is used to separate the fluid particles into two groups, as explained in section~\ref{sec:methods} above.   The group containing the reference particle before the last binary combination of the hierarchical clustering comprises particles inside the same coherent structure as the reference particle.  The results of this clustering are shown in figure~\ref{fig:gyre4}.  It is clear that the thresholding algorithm using multi-dimensional eigen-distance from a specified reference point is able to isolate the points inside the upper left vortex, and its boundary aligns with a region of locally high gradients in the CSC field (figure~\ref{fig:gyre1}(a)), as well as the gyre core identified in the FTLE field using 65,000 particles (figure~\ref{fig:gyre4}(b)).

\begin{figure}
\centering
\includegraphics[width=\linewidth]{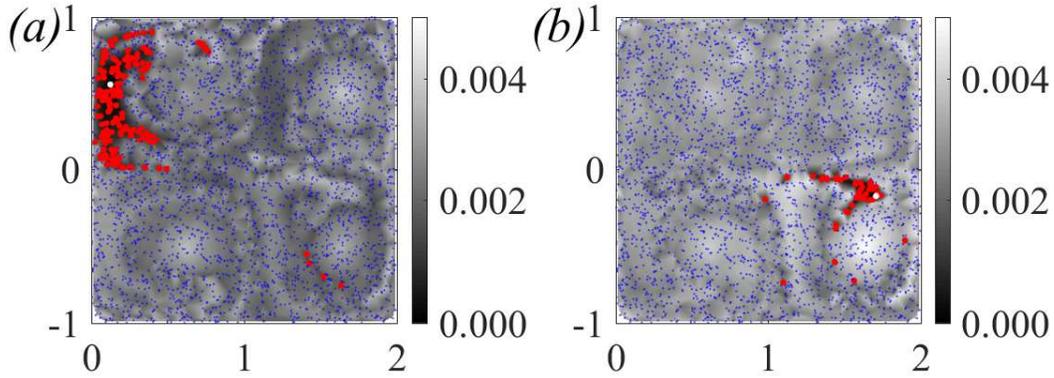}
\caption{a) Analysis of a reference particle to the left of the vortex core of the upper left quadrant of the quadruple gyre flow.  The underlying field is the 9D unweighted distance metric field. b) Analysis of a reference particle above the vortex core of the lower right quadrant of the quadruple gyre flow.  The underlying field is the 8D unweighted distance metric field.  Dots indicate final particle positions: white dot indicates the chosen reference point, red dots are the particles that have been identified as within the coherent structure associated with the reference point, and blue dots  indicate particles outside of that structure.  }
\label{fig:gyre5}
\end{figure}

This process can be applied to any choice of reference trajectory.  Generally, reference points of interest will be local maxima or minima in the CSC field, but this is not always the case.  For example, from the CSC field, it is clear that there are relatively large regions of the flow containing particles with a relatively constant value of coloring near $\textrm{CSC}=2.5\times10^{-4}$ (i.e. yellow regions in figure~\ref{fig:gyre1}(a)).  These regions are concentrated to the left of the gyre cores.  To investigate the presence of large scale structures corresponding to these particles, we choose a reference particle in the middle of this region in the upper left quadrant, indicated by the white dot in figure~\ref{fig:gyre5}(a).  
The plot of mean slope of the distance metric vs. dimensionality of the eigenspace (not shown), indicates that the eigenspace with dimensionality nine should be sufficient for ensuring a plateau in the distance-metric field.  Using the CSC field and subsequent 8 eigenvectors, the unweighted distance-metric field for 9-dimensional eigenspace is calculated, and hierarchical clustering is used to identify particles in the same coherent structure as the reference particle, as seen in figure~\ref{fig:gyre5}(a).  It is significant that the particles inside this structure are consistent with the boundary of the corresponding structure in the double gyre flow analysis of~\citet{Allshouse2015} using the fuzzy c-means  clustering algorithm, where it was necessary to prescribe the total number of coherent structures apriori.

Another point of interest in this CSC field is one with a large negative CSC value.  A point above the bottom right gyre core was selected, and is indicated by the white dot in figure~\ref{fig:gyre5}(b).  As shown in a previous analyis~\citep{Schlueter-Kuck2017}, this particle and others with large negative CSC values have trajectories that are characterized by switching quadrants in the time domain of interest.  Using the aforementioned techniques, the eight-dimensional eigenspace is found to be appropriate for examination of this structure.  The hierarchical clustering approach identifies a connected region of particles above the lower right gyre core with a few particles that are not connected (but still considered coherent), shown by red dots in figure~\ref{fig:gyre5}(b).

\subsection{Bickley Jet}\label{sec:bickleyjet}
The Bickley jet, another analytical example, is frequently used as a model of zonal jets in the Earth's atmosphere~\citep{Rypina2007}.  It is a quasi-periodic flow comprising a spatially undulatory jet with counter-rotating vortices above and below.  Here we use the Bickley jet to show the success of this method for more complex flows and lower particle densities.  The flow is described by the stream function $\psi=\psi_0+\psi_1$, where
\begin{eqnarray}
\label{eq:bickley_eqn1a}
\psi_0 & =&c_3y-UL\tanh\left(y/L\right)\\
\label{eq:bickley_eqn1b}
\psi_1 & =&UL\textrm{ sech}^2\left(y/L\right)\sum_{n=1}^3\epsilon_n\cos\left(k_n\left(x-\sigma_nt\right)\right)
\end{eqnarray}

We use similar values of the parameters as in~\citet{Hadjighasem2016}: $U=62.66$ ms$^{-1}$, $L=1770$ km, $k_n=2n/r_0$, $c=[0.1446U$, $0.205U$, $0.461U]$,  $\sigma=c-c(3)$, and $\epsilon=[0.0075$, $0.15$, $0.3]$, and the flow is computed on the interval $x=[0$, $20\times10^6]$ m, $y=[-3\times10^6$, $3\times10^6]$ m, over the time interval $t=[0$, $40]$ days, divided into 601 discrete time steps.  The flow was considered periodic in $x$.  For calculation of the CSC, 300 particles were initialized randomly in the domain and advected with the flow.  The particles were followed over the entire time interval, even if they left the domain, analogous to how ocean drifters are tracked.

\begin{figure*}
\centering
\includegraphics[width=\linewidth]{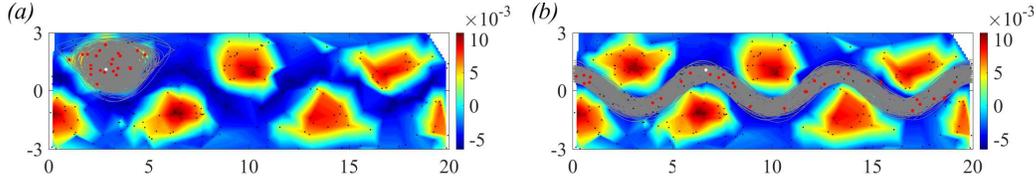}
\caption{a) Analysis of a reference particle in the vortex core of the upper left vortex of the Bickley jet flow.  b) Analysis of a reference particle in the jet.  Dots indicate final particle positions: white dot indicates the chosen reference point, red dots are the particles that have been identified as within the coherent structure associated with the reference point, and black dots  indicate particles outside of that structure.  The underlying field is the CSC field calculated using 300 particles. Gray lines indicate the full trajectories of the particles inside the coherent structure associated with the reference point (white and red particles)} 
\label{fig:bickley6}
\end{figure*}
As with the quadruple gyre, the CSC field and subsequent eigenvectors can be used to identify coherent structures associated with individual particles in the flow field.  The CSC field for the Bickley jet seeded with 300 particles can be found in~\citet{Schlueter-Kuck2017}.  The CSC field identifies a meandering jet flanked above and below by counter rotating vortices.  Several reference points in the flow were analyzed, and the resulting coherent structures associated with this points are explained and  visualized here.  The first reference point of interest is located near the local maximum in the CSC field in the counterclockwise rotating vortex in the upper left of the domain, as is indicated by the white dot in figure~\ref{fig:bickley6}(a).  Using analysis similar to that performed for the quadgyre, we determined that the six eigenvectors associated with the largest six eigenvalues of the generalized eigenvalue problem were sufficient for analysis in this  case, using the plot of mean slope vs. dimensionality of the eigenspace (not shown).  Using hierarchical clustering to group the six-dimensional eigenspace into two clusters, the upper left vortex core was isolated from the rest of the flow.  This is illustrated in figure~\ref{fig:bickley6}(a), where the reference point is white, the other particles contained in the coherent structure with the reference point are in red, and the points outside of the structure are black.  The dots are overlaid on the CSC field for the Bickley jet calculated using 300 tracer particles for comparison.  Also shown in this figure are the full trajectories of the particles identified by this algorithm as within the structure associated with the reference point, given by the gray lines.  It is evident that all of these trajectories remain tightly entwined throughout the time domain.  Several black particles appear to be within the domain encompassed by the trajectories; this is due to the unsteadiness of the flow which causes the vortex to oscillate slightly left to right in the chosen reference frame.  
\begin{figure*}
\centering
\includegraphics[width=\linewidth]{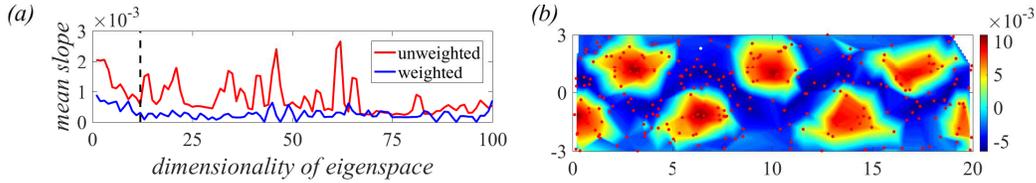}
\caption{Analysis of a reference particle in the background flow.  a) the average slope of the distance to the reference particle ($s(D)$) vs. the dimensionality of the eigenspace ($D$).  Red line is calculated using the unweighted distance metric, where $w_d=1$.  Blue line is calculated using the weighted distance metric, where $w_d$ is given by equation~\ref{eq:dist_weight}.  Vertical dotted black line indicates the location of the elbow in the plot, selected for subsequent analysis ($D=12$).  b) Dots indicate final particle positions: white dot indicates the chosen reference point, red dots are the particles that have been identified as within the coherent structure associated with the reference point, and black dots  indicate particles outside of that structure.  The underlying field is the CSC field calculated using 300 particles.} 
\label{fig:bickley7}
\end{figure*}

The same algorithm can be applied to a reference point in the jet, as shown in figure~\ref{fig:bickley6}(b).  In this case, the distance to the reference point in three-dimensional eigenspace was examined.  It is evident from figure~\ref{fig:bickley6}(b) that this algorithm clearly identifies the jet as a coherent structure in the flow.  This is significant because the jet does not fulfill many of the requirements of a coherent structure under other traditional analysis techniques; it is not convex or nearly convex, and because the jet spans the full domain of the flow, many of the particles remain very far apart from each other.  However, over the time interval of the flow under analysis, the particles in the jet remain distinct from the particles in the surrounding fluid.  As a result, this structure could potentially be very good at transporting scalar quantities.  The Coherent Structure Coloring algorithm admits the jet as a coherent structure even though other algorithms would miss it.

Finally, structure identification can be used to analyze a reference point in the background flow.  The results of this investigation are shown in figure~\ref{fig:bickley7}(b), in which a 12-dimensional eigenspace distance was considered (see figure~\ref{fig:bickley7}(a)).  In this case, the hierarchical clustering algorithm identified 298 of the 299 additional particles (in red) as belonging to a coherent structure associated with the reference point (in white).  There are several indications that the reference particle in this case is not part of any coherent structure.  First, figure~\ref{fig:bickley6}(a) and (b) have identified structures that do not contain the current reference point in the background flow, while the analysis of the particle in the background flow admits almost all particles into the structure associated with it.  True coherency should be commutative.  Additionally, a comparison of the mean slope versus dimensionality of the eigenspace for a reference particle in the upper left vortex of the quadgyre flow (shown in figure~\ref{fig:gyre3}), and a reference point in the background flow of the Bickley jet (shown in figure~\ref{fig:bickley7}(a)) provides other indications.  For the quadgyre vortex analysis, the ``elbow" in this plot is apparent, and it is straightforward to determine the dimensionality of the eigenspace that is appropriate for analysis (6D).  However, for the particle in the background flow, there is no clear limit after which subsequent eigenvectors can be ignored, even up to 100-dimensional eigenspace.  

\section{Conclusions}\label{sec:conclusions}
Through the use of two analytical flows, we have shown an effective method for extracting coherent structures by using the CSC field and subsequent eigenvectors and eigenvalues of the generalized eigenvalue problem $LX=\lambda DX$.  This method involves selecting a Lagrangian particle trajectory of interest in the flow and calculating a weighted distance metric in multi-dimensional eigenspace to threshold the flow (e.g. using agglomerative hierarchical clustering).  These threshold values are shown to effectively separate coherent structures from the background flow, and the boundaries of these structures are consistent with regions where the gradients in the CSC field are high.  This method is shown to be robust to the size and shape of the structures being identified, and allows for a more versatile definition of coherence then is generally allowed by other methods.  The CSC field  can inform reference point selection, with regions of local maxima and minima in the CSC field yielding good candidates for flow trajectories associated with dominant flow structures.  

It is important to emphasize that this approach is effective for identifying structures associated with individual flow trajectories, but is not intended to identify the full set of coherent structures in a flow, as is a common goal with other clustering and spectral graph theory methods~\citep{Hadjighasem2016,Froyland2015,Froyland2017}.  Application of the method described here to the full set of Lagrangian trajectories could potentially be used to identify the full set of coherent structures associated with those trajectories.  The anticipated commutative property of the coherency relationships might be leveraged to achieve efficient computation of coherent sets.

Although the method is applied here to cluster tracer particles in fluid flows, the generality of the approach allows for its potential application to other unsupervised clustering problems in dynamical systems such as neuronal activity, gene expression, or social networks.  In these latter cases, the only difference from the present application is that the current kinematic trajectory description (i.e. tracer position versus time) is replaced with other descriptions that facilitate analogous calculations of dissimilarity among the trajectories, e.g. action potential versus time for a collection of neurons; expression versus gene locus in a microarray; or agent behavior in a social community.

A MATLAB implementation of the CSC algorithm is available for free download at http://dabirilab.com/software.

\begin{acknowledgments}
This work was supported by the U.S National Science Foundation and by the Department of Defense (DoD) through the National Defense Science $\&$ Engineering Graduate Fellowship (NDSEG) Program.
\end{acknowledgments}

\nocite{*}

\end{document}